\documentclass[a4paper,12pt]{article}
\usepackage{amsmath,amsfonts,amssymb}

\begin{document}

\begin{center}
{\Large\bf Note on relation between bottom-up holographic models
and large-$N_c$ QCD}
\end{center}

\begin{center}
{\large S. S. Afonin}
\end{center}

\begin{center}
{\it Saint Petersburg State University, 7/9 Universitetskaya nab.,
St.Petersburg, 199034, Russia}
\end{center}

\begin{abstract}
We discuss a derivation of the quadratic in fields part of action
of bottom-up holographic models from some general properties of
the large-$N_c$ limit in QCD. 
\end{abstract}

\section{Introduction}

The physics of hadrons composed of light quarks is highly
non-perturbative as these hadrons represent strongly coupled
systems. Some time ago an interesting proposal of analytical
approach to the strongly coupled gauge theories came from the
string theory --- the hypothesis of AdS/CFT
correspondence~\cite{mald} (also referred to as gauge/gravity
duality or holographic duality). It turns out that a description
of certain strongly coupled gauge theories can be given in terms
of weakly-coupled higher-dimensional gravitational theories. The
holographic ideas have penetrated to many branches of modern
theoretical physics. In the hadron spectroscopy, these ideas are
challenging since the AdS/CFT correspondence provides a practical
recipe for calculation of correlation functions in strongly
coupled gauge theories. The calculation is performed via the
semiclassical expansion of the action of higher-dimensional dual
theories~\cite{witten}. This can have a direct application to the
hadron spectroscopy: The masses of hadron states appear as poles
of correlation functions of QCD currents interpolating those
states.

The concrete holographic prescription is as follows~\cite{witten}.
All correlation functions of a 4D field theory can be obtained
from the generating functional of the connected correlators
$W_{4D}[\varphi_0(x)]$ that depends on the sources $\varphi_0(x)$
for the 4D field theory operators. If we know the action $S_{5D}$
of dual theory, the holographic correspondence postulates the
identification
\begin{equation}
\label{4a}
W_{4D}[\varphi_0(x)]=S_{5D}[\varphi(x,\epsilon)].
\end{equation}
By assumption, the 5D dual theory is in the weakly coupled regime.
This implies a important consequence: As the first approximation,
one may consider only the quadratic terms in $S_{5D}$ and use
the semiclassical analysis. The calculation of $n$-point correlators
consists thus in evaluation of $S_{5D}$
on a solution of equation of motion and subsequent
differentiation $n$ times with respect to boundary values of 5D fields.

The bottom-up holographic approach represents an attempt to apply the
holographic ideas to QCD. The corresponding models try to describe the
non-perturbative dynamics of strong interactions in terms of a
putative semiclassical five-dimensional theory. The first versions
of bottom-up models --- the so-called Hard Wall (HW) models~\cite{son1,pom}
--- had a problem with description of linear Regge and radial spectrum
($m_n\sim n$ instead of the behavior
$m_n^2\sim n$ expected in the phenomenology and on some theoretical grounds).
The construction of the Soft Wall (SW)
holographic model~\cite{son2} solved this problem.

There is still no consensus why the bottom-up holographic models
turned out to be unexpectedly successful in the phenomenology.
This success looks surprising and a bit enigmatic. An inherent
property of the holographic duality is the large-$N_c$ (planar)
limit of gauge theories~\cite{hoof}. It is believed therefore that
all AdS/QCD models are models for the planar QCD where mesons
represent stable and freely propagating particles (their
interactions and decays are suppressed by powers of
$1/N_c$~\cite{hoof}). Within a hypothetical dual theory, the
corresponding mass spectrum should emerge in the weak coupling
regime, i.e., in the first approximation, it must be determined by
a quadratic in fields part of action.

The purpose of this work is to demonstrate explicitly that the
quadratic in fields parts of actions of typical bottom-up
holographic models follow directly from some general expectations
about the behavior of correlation functions in the large-$N_c$
limit of QCD. Partly this analysis was done in
Ref.~\cite{afoninI}. But the derivation in~\cite{afoninI} did not
take into account that, in general case, one must use rescaled
(with respect to the holographic coordinate) fields in order to
have finite sources. In the
present work, we fix this drawback. As a byproduct, we also
demonstrate the importance of "weak gravity" approximation which
allows to replace the covariant derivatives by the usual ones and
to use a important relation for the 5D mass as a function of spin.

\section{The large-$N_c$ QCD as a Kaluza-Klein reduction}

We recall some general results from the analysis of QCD in the limit
of large number of colors $N_c$~\cite{hoof}. The meson masses scale
as $m\sim N_c^0$ and their full decay widths do as $\Gamma\sim
N_c^{-1}$.  The decay widths thus disappear in the limit of large $N_c$.
Since the meson masses vary slightly with $N_c$, this limit is
often useful in the meson spectroscopy.
In the extreme case
$N_c\rightarrow\infty$, the mesons are infinitely narrow and
non-interacting, in addition, an infinite number of states
emerge for each set of quantum numbers. Such an infinite
tower of resonance poles saturates completely the two-point
correlation function of quark currents with quantum numbers
$I^G(J^{PC})$ corresponding to the given tower,
\begin{equation}
\label{1e}
\left\langle\mathcal{O}_J(q)\mathcal{O}_J(-q)\right\rangle\sim\sum_{n=0}^{\infty}\frac{(F_n^{(J)})^2}{q^2-m_{n,J}^2+i\varepsilon},
\end{equation}
With the help of functional integral formalism, this is described
by the following action with source terms,
\begin{equation}
\label{1}
I_{[\mathcal{O}_J]}=(-1)^{J}\int d^4x\sum_{n=0}^{\infty}\left(\partial_{\mu}\phi_J^{(n)}\partial^{\mu}\phi^J_{(n)}-
m_{n,J}^2\phi_J^{(n)}\phi^J_{(n)}+\dots+\phi^J_{(n)}\mathcal{O}_J^{(n)}\right),
\end{equation}
Here $\phi_J\doteq\phi_{\mu_1\mu_2\dots\mu_J}$, $\mu_i=0,1,2,3$
(the sign convention is $(+---)$), corresponds to a meson field of spin
$J$, the other quantum numbers are unspecified. The tensor $\phi_J$ is
symmetric, traceless, $\phi^{\mu}_{\mu\dots}=0$ (to provide the
irreducible $\left(J/2,J/2\right)$ representation of the
homogeneous Lorentz group), and satisfy the auxiliary condition
$\partial^{\mu}\phi_{\mu\dots}=0$ (to give the required $2J+1$
physical degrees of freedom). The dots denote additional
derivative terms appearing in description of free higher spin
fields. The auxiliary conditions can be chosen in such a way that they do
not contribute (to be discussed below). The tensor structure in
the r.h.s. of representation~\eqref{1e} depends on the structure
of derivative terms. The sources $\mathcal{O}_J^{(n)}$ can be
represented as
\begin{equation}
\label{1b}
\mathcal{O}_J^{(n)}=F_n^{(J)}\mathcal{O}_J,
\end{equation}
where the constants $F_n^{(J)}$ are defined by
\begin{equation}
\label{1c}
\langle 0|\mathcal{O}_J^{(n)}|\phi_J^{(n)} \rangle=F_n^{(J)}\varepsilon_J
\end{equation}
for a meson $\phi_J^{(n)}$ with "polarization" $\varepsilon_J$ and
$\mathcal{O}_J$ is a common source to which the states
$\phi^J_{(n)}$ are coupled with a "coupling" $F_n^{(J)}$. The
representation~\eqref{1b} follows from the requirement to get the
standard form for two-point correlators: integrating over the
field $\phi_J$ in the generating functional
\begin{equation}
\label{1d}
Z[\mathcal{O}_J]=\int D \phi_J e^{I_{[\mathcal{O}_J]}},
\end{equation}
and differentiating twice with respect to $\mathcal{O}_J$ at
$\mathcal{O}_J=0$ one arrives at the sum over meson poles~\eqref{1e}.

Our task is to rewrite the expression~\eqref{1} as a Kaluza-Klein reduction
of action of some free 5D theory. We will look for the corresponding 5D actions in the form
\begin{equation}
\label{2}
S_{\text{5D}}=(-1)^{J}\int d^4x\,dz f_1(z)\left(\partial_M\varphi_J\partial^M\varphi^J-
m_J^2f_2(z)\varphi_J\varphi^J\right),
\end{equation}
where $M=0,1,2,3,4$, $\varphi_J=\varphi_{M_1M_2\dots M_J}(x,z)$, and $f_1(z)$,
$f_2(z)$ are yet unknown functions of the fifth (space-like)
coordinate $z$. We start from a flat 5D space having boundaries
in the fifth coordinate, $z_{\text{min}}\leq z \leq z_{\text{max}}$.
The action~\eqref{2} contains a $z$-dependent background which
is different for the kinetic and mass terms.

We will imply the following condition for the physical components
of 5D fields,
\begin{equation}
\label{8}
\varphi_{z\dots}=0.
\end{equation}
With this condition, no additional
quadratic terms appear in description of free higher-spin mesons~\cite{son2,katz,br2}.

The equation of motion for the action~\eqref{2} with condition~\eqref{8} is
\begin{equation}
\label{9}
-\partial_z[f_1(z)\partial_z\varphi_n^{(J)}(z)]+f_1(z)f_2(z)m_J^2\varphi_n^{(J)}(z)=m_{n,J}^2f_1(z)\varphi_n^{(J)}(z),
\end{equation}
where we used the usual plane-wave ansatz for the physical
4D particles carrying the 4D momentum $q_n$,
\begin{equation}
\label{11}
\varphi(x,z)=e^{iq_nx}\varphi_n(z),\qquad q_n^2=m_n^2.
\end{equation}
After imposing boundary conditions on the fields, the equation~\eqref{9}
represents a classical Sturm-Liouville (SL) problem.
Consider the following mathematical problem: Given a
spectrum $m_{n,J}^2$ in~\eqref{1}, find the functions
$f_1(z)$ and $f_2(z)$ which being inserted in~\eqref{9} would lead to
eigenvalues $m_{n,J}^2$. The solution of this "inverse" SL problem constitutes the
main intermediate step in rewriting the action~\eqref{1} with sources
as some free 5D field theory.

First we recall briefly the main results of the SL theory. The SL equation is
\begin{equation}
\label{a1}
-\partial_z[p(z)\partial_z\varphi]+q(z)\varphi=\lambda\omega(z)\varphi.
\end{equation}
Here the function $p(z)>0$ has a continuous derivative, the
functions $q(z)>0$ and $\omega(z)>0$ are continuous on the finite
closed interval $[z_{\text{min}},z_{\text{max}}]$.
The SL problem consists in finding the values of $\lambda$ for which
there exists a non-trivial solution of Eq.~\eqref{a1} satisfying certain
boundary conditions. Under the assumptions that $p(z)^{-1}$, $q(z)$,
and $\omega(z)$ are real-valued integrable functions over the
interval $[z_{\text{min}},z_{\text{max}}]$, with the boundary
conditions of the form
\begin{equation}
\label{a2}
\varphi(z_{\text{min}})\cos\alpha-p(z_{\text{min}})\varphi'(z_{\text{min}})\sin\alpha=0,
\end{equation}
\begin{equation}
\label{a3}
\varphi(z_{\text{max}})\cos\beta-p(z_{\text{max}})\varphi'(z_{\text{max}})\sin\beta=0,
\end{equation}
where $\alpha,\beta\in[0,\pi)$ and prime means derivative, the SL theorem states that
(i) There is an infinite discrete set of real eigenvalues $\lambda_n$, $n=0,1,2,\dots$;
(ii) Up to a normalization constant, there is a unique eigenfunction $\varphi_n(z)$
corresponding to each eigenvalue $\lambda_n$ and this eigenfunction has exactly $n-1$
zeros in $[z_{\text{min}},z_{\text{max}}]$;
(iii) The normalized eigenfunctions form an orthonormal basis
\begin{equation}
\label{a4}
\int_{z_{\text{min}}}^{z_{\text{max}}}\varphi_m(z)\varphi_n(z)\omega(z)dz=\delta_{mn}.
\end{equation}
Thus the solutions of the SL problem form a complete set of
functions in the interval $[z_{\text{min}},z_{\text{max}}]$ which
can be used for expansion of arbitrary (but sufficiently smooth) functions in that
interval.

Under these conditions, the SL problem~\eqref{9} has solutions $\varphi_n^{(J)}(z)$
which are normalized as
\begin{equation}
\label{12}
\int_{z_{\text{min}}}^{z_{\text{max}}}f_1(z)\varphi_m^{(J)}(z)\varphi_n^{(J)}(z)dz=\delta_{mn},
\end{equation}
and form a complete set of functions. The function $\varphi_J(x,z)$ in~\eqref{2} can be expanded
in the 4D harmonics,
\begin{equation}
\label{13}
\varphi_J(x,z)=\sum_{n=0}^{\infty}\phi_J^{(n)}(x)\varphi_n^{(J)}(z).
\end{equation}
Now we substitute the expansion~\eqref{13} into the action~\eqref{2},
\begin{multline}
S_{\text{5D}}=(-1)^{J}\int d^4x\,dz f_1(z) \sum_{m,n=0}^{\infty} \left\{\varphi_m^{(J)}\varphi_n^{(J)}
\partial_{\mu}\phi_J^{(m)}\partial_{\mu}\phi_J^{(n)}\right.\\
\left.-\phi_J^{(m)}\phi_J^{(n)}\partial_z\varphi_m^{(J)}\partial_z\varphi_n^{(J)}
-m_J^2f_2(z)\varphi_m^{(J)}\varphi_n^{(J)}\phi_J^{(m)}\phi_J^{(n)}\right\}.
\label{14}
\end{multline}
Integrating by parts
and making use of Eq.~\eqref{9}, the second term in the
action~\eqref{14} can be rewritten as (dropping the general factor
$-\phi_J^{(m)}\phi_J^{(n)}$)
\begin{multline}
\int_{z_{\text{min}}}^{z_{\text{max}}} dz f_1(z)\partial_z\varphi_m^{(J)}\partial_z\varphi_n^{(J)}=\\
\left.\varphi_m^{(J)}f_1(z)\partial_z\varphi_n^{(J)}\right\vert_{z_{\text{min}}}^{z_{\text{max}}}+
\int_{z_{\text{min}}}^{z_{\text{max}}} dz \varphi_m^{(J)}\varphi_n^{(J)}g_J(z)\left(m_{n,J}^2-f_2(z)m_J^2\right).
\label{15}
\end{multline}
Integrating over $z$ in the action~\eqref{14} with
the help of~\eqref{12} and~\eqref{15} we obtain
\begin{multline}
S_{\text{5D}}=(-1)^{J}\int d^4x\sum_{n=0}^{\infty}\left\{\left(\partial^{\mu}\phi^J_{(n)}\right)^2-
m_{n,J}^2\left(\phi^J_{(n)}\right)^2\right.\\
\left.-\left.\phi^J_{(n)}f_1(z)\partial_z\varphi_n^{(J)}\varphi_J(x,z)\right\vert_{z_{\text{min}}}^{z_{\text{max}}}\right\},
\label{16}
\end{multline}
with $\varphi_J(x,z)$ given by~\eqref{13}.
The action~\eqref{16} has the form of~\eqref{1} if we identify
\begin{equation}\label{17}
\mathcal{O}_J^{(n)}=-\left.f_1(z)\partial_z\varphi_n^{(J)}\varphi_J(x,z)\right\vert_{z_{\text{min}}}^{z_{\text{max}}}.
\end{equation}

The classical Kaluza-Klein reduction corresponds to $\mathcal{O}_J^{(n)}=0$.
But the case $\mathcal{O}_J^{(n)}\neq0$ is possible as well. And this
very possibility turns out to be equivalent to the holographic prescription for
obtaining the 4D Green functions.

Choosing $\beta=0$ or $\beta=\pi/2$ in the boundary condition~\eqref{a3}
we can nullify the term at $z=z_{\text{max}}$ in Eq.~\eqref{17}.
Note in passing that for the vector mesons the first
possibility is actually realized in the SW holographic
models~\cite{son2}, where $z_{\text{max}}\rightarrow\infty$, while
the second one --- in the HW models~\cite{son1,pom},
where $z_{\text{max}}$ is the infrared cutoff. Thus,
\begin{equation}
\label{18}
\mathcal{O}_J^{(n)}=\lim_{z\rightarrow z_{\text{min}}+0} f_1(z)\partial_z\varphi_n^{(J)}(z)\varphi_J(x,z).
\end{equation}

\section{From large-$N_c$ QCD to bottom-up holographic models}

The action~\eqref{2} can be rewritten as a "weakly coupled"
5D gravitational theory of free field $\varphi_J$ with the metric
\begin{equation}
\label{4}
ds^2=f_2(z)(dx_{\mu}dx^{\mu}-dz^2),
\end{equation}
that dictates the corresponding metric tensor $G_{MN}$ and the background
function $f_1(z)$,
\begin{equation}
\label{3}
f_1(z)=\sqrt{|\det{G_{MN}}|}=f_2^{\frac52}(z).
\end{equation}
We must impose the covariant rule for contraction of indices,
\begin{equation}
\label{5}
\partial_M\varphi_J\partial^M\varphi^J=\partial_M\varphi_{M_1\dots M_J}\partial_{M'}\varphi_{M_1'\dots M_J'}
G^{MM'}G^{M_1M_1'}\dots G^{M_JM_J'},
\end{equation}
where
\begin{equation}
\label{6}
G^{MN}=G^{-1}_{MN}=f_2^{-1}(z)\eta^{MN}.
\end{equation}
This can be achieved only if the function $f_1(z)$ in~\eqref{3}
is replaced by the $J$-dependent background
\begin{equation}
\label{3b}
f_1^{(J)}(z)=f_2^{\frac32-J}(z).
\end{equation}
All this just rearranges the notations but does not change our discussions above.

The choice
\begin{equation}
\label{4b}
f_2(z)=\frac{R^2}{z^2}, \qquad 0\leq z<\infty,
\end{equation}
in~\eqref{4} corresponds to the AdS space of radius $R$.
This choice will be exploited in what follows to find connections
with the holographic approach.

Now we should identify the general source $\mathcal{O}_J$. The identification
must provide finite and non-zero values for constants $F_n^{(J)}$ in the representation~\eqref{1b}.
Our prescription is
\begin{equation}
\label{19b}
\mathcal{O}_J=zf_1^{(J)}\varphi_J/c\sim z^{2(J-1)}\varphi_J/c,
\end{equation}
\begin{equation}
\label{19}
F_n^{(J)}=\lim_{z\rightarrow +0}c\frac{\partial_z\varphi_n^{(J)}(z)}{z}.
\end{equation}
The eigenfunctions $\varphi_n^{(J)}(z)$ in the AdS space
behave as $z^2$ at small $z$ for the twist 2 operators,
i.e. when the relation~\eqref{b5D} holds~\cite{afoninI}.
This ensures constant values
for $F_n^{(J)}$ in~\eqref{19}. The appearance of
factor $c$ is due to a freedom to redefine
$F_n\rightarrow cF_n$, $\mathcal{O}\rightarrow \mathcal{O}/c$ where
the parameter $c$ can be fixed by matching to the high energy
asymptotics of the corresponding QCD correlators. For
the case of vector mesons,
our way of getting the relation~\eqref{19} is alternative
(up to a general factor depending on normalization of fields)
to the derivation of Ref.~\cite{son1}.

The relation~\eqref{19b} shows that if we want, following the AdS/CFT prescriptions,
to interpret the source as a finite boundary value of a 5D field,
$\varphi_J(x,+0)=\mathcal{O}_J$, we must rescale
the higher spin ($J>1$) fields,
\begin{equation}
\label{19c}
\varphi_J=z^{2(1-J)}\tilde{\varphi}_J.
\end{equation}
In this sense, the physical tensor field is the rescaled one $\tilde{\varphi}_J$.
This statement was obtained in Ref.~\cite{katz}
using a different approach.

Now we see that the holographic prescription for calculation of correlation functions in a 4D
theory from a "dual" 5D theory is completely reconstructed on the level of quadratic in
fields actions. The higher vertices (describing meson decays and scattering) are suppressed
in the large-$N_c$ limit and we therefore do not consider them.

Some clarifying remarks should be made. Since the AdS space is not flat, the usual
derivatives must be replaced by the covariant ones in the case of tensor mesons.
However, the holographic duality for strongly coupled 4D theories
works only if the dual 5D theories are weakly coupled. This means that
we should use the "weak gravity" limit (very large AdS radius) as the first approximation. In this
limit, the affine connections are neglected as the AdS space locally is almost flat. This justifies the use of
usual derivatives for higher spin mesons in our analysis.

The second remark concerns the standard statement that the
spectrum of the HW model depends on the boundary conditions.
Indeed, the Dirichlet boundary conditions which have been used
since the pioneering papers~\cite{son1,pom} correspond to the
choice $\beta=\pi/2$ in~\eqref{a3} and seemingly one could use any
other choice for $\beta$. There is, however, a subtle point here.
With other choices, generally speaking, the boundary contribution
at $z=z_{\text{max}}$ is not zero. This means the appearance of
the second source that would look strange for the standard
functional technics\footnote{The existence of two sources is not
necessarily senseless in the real strong interactions where the
physical degrees of freedom at low and high energies are different
(pseudogoldstone bosons {\it vs.} quarks and gluons). For
instance, the quark vector current $\bar{q}\vec{\tau}\gamma_\mu q$
interpolates the $\rho$-meson at high enough energies where the
current quarks are well defined. At low enough energies, the
$\rho$-meson could be considered as a two-pion state. 
The ultraviolet and infrared sources should be then different as long
as the physical degrees of freedom are different. As far as we
know, this possibility was not exploited in the bottom-up
approach.}.

The last remark concerns the choice of background in the 5D
action. It can be shown that if we want to reproduce the linear
Regge spectrum and simultaneously the correct analytical structure
of the operator product expansion of the two-point correlators,
the background is fixed unambiguously --- it corresponds to
nothing but the SW model to be discussed in the next section. The
corresponding analysis is presented in detail in
Ref.~\cite{afoninI}.

\section{Spectrum of the soft-wall model}

An attractive feature of the SW holographic models is the Regge spectrum for meson masses~\cite{son2}.
The original derivation of Ref.~\cite{son2} was based on gauge higher spin fields. The
condition~\eqref{8} was interpreted as a gauge choice. A residual invariance after imposing~\eqref{8}
(that canceled the additional derivative terms) allowed to eliminate the mass term for the rescaled
fields~\eqref{19c}. Below we reproduce the Regge spectrum of Ref.~\cite{son2} using the
massive higher spin fields and our discussions above.

The action of SW model is a particular case of~\eqref{2} written in the AdS space,
\begin{equation}
\label{b2}
S_{\text{5D}}=(-1)^{J}\int d^4x\,dz \sqrt{|\det{G}|}\,e^{az^2}\left(\partial_M\varphi_J\partial^M\varphi^J-
m_J^2\varphi_J\varphi^J\right),
\end{equation}
where the dilaton background $e^{az^2}$ was introduced to get the Regge spectrum.
Substituting the rescaled field~\eqref{19c} and lowering all indices we find
\begin{multline}
\label{c2}
S_{\text{5D}}=(-1)^{J}R^{2J-1}\int d^4x\,dz e^{az^2}\left\{\left(\partial_M\tilde{\varphi}_J\right)^2z^{1-2J}\right.\\
\left.-\left((2-2J)^2+m_J^2R^2\right)z^{-1-2J}\tilde{\varphi}_J^2
-4(1-J)z^{-2J}\tilde{\varphi}_J\partial_z\tilde{\varphi}_J\right\}.
\end{multline}
The corresponding equation of motion for physical modes~\eqref{11} reads
\begin{multline}
\label{b9}
-\partial_z\left[\frac{e^{az^2}}{z^{2J-1}}\partial_z\tilde{\varphi}_{n,J}\right]+
\frac{m_J^2R^2-4(J-1)}{z^{2J+1}}e^{az^2}\tilde{\varphi}_{n,J}+
\frac{4a(1-J)}{z^{2J-1}}e^{az^2}\tilde{\varphi}_{n,J}\\
=m_{n,J}^2\frac{e^{az^2}}{z^{2J-1}}\tilde{\varphi}_{n,J}.
\end{multline}
The SL equation~\eqref{b9} has the following discrete spectrum
\begin{equation}
\label{spec}
m_{n,J}^2=2|a|\left[2n+1+\frac{a}{|a|}(J-1)+\sqrt{(J-2)^2+m_J^2R^2}\right].
\end{equation}
This spectrum depends on the 5D mass $m_J$. The free $p$-form fields in the AdS$_5$ space
have masses $m_p^2R^2=(\Delta-p)(\Delta+p-4)$. As was shown in Ref.~\cite{br2},
the condition~\eqref{8} leads to decoupling of kinematical aspects and the
symmetric tensors of rank $p$ (describing mesons of spin $p$) obey the same
equation as the $p$-form fields. In the "weak gravity" approximation,
this results in the following 5D mass of higher spin fields
propagating in the AdS$_5$ space~\cite{br2,gutsche},
\begin{equation}
\label{5D}
m_J^2R^2=(\Delta-J)(\Delta+J-4),
\end{equation}
where $\Delta$ is the canonical (conformal) dimension of a 4D
field theory operator dual to a corresponding 5D field on the AdS boundary.

The spectrum of the original SW model~\cite{son2} corresponds
to the operators of twist 2, i.e. $\Delta=J+2$. Substituting this
dimension to~\eqref{5D}, we have
\begin{equation}
\label{b5D}
\left.m_J^2R^2\right|_{{}_{\Delta=J+2}}=4(J-1),
\end{equation}
The spectrum~\eqref{spec} reduces then to the spectrum,
\begin{equation}
\label{spec2}
m_{n,J}^2=2|a|\left[2n+J+1+\frac{a}{|a|}(J-1)\right].
\end{equation}
This spectrum does not depend on spin if $a<0$.
For $a>0$ one arrives at a remarkably simple Regge spectrum
$m_{n,J}^2=4a(n+J)$ that coincides with the spectrum of poles of the Veneziano amplitude.


For the purposes of phenomenological applications, it can be interesting
to construct an extension of the SW model that leads to a
shifted Regge spectrum, $m_{n,J}^2=4a(n+J+b)$, where $b$ is
a phenomenological parameter regulating the intercept of
Regge and radial trajectories. The matter is that this form
of light meson spectrum (with $b\neq0$) seems to be supported by
the experimental data~\cite{lin}. The relevant holographic extension was constructed
in Ref.~\cite{genSW} for the vector case. The method of
Ref.~\cite{genSW} can be generalized to higher spins.
We will not display the derivation in detail
and just give the final answer that can be checked by a direct
calculation: The dilaton background $e^{az^2}$ must be replaced by
\begin{equation}
e^{az^2}\rightarrow e^{az^2}U^2(b,-|J-1|;az^2),
\end{equation}
where $U$ is the Tricomi hypergeometric function.

\section{In conclusion}

In essence, we tried to demonstrate that the bottom-up holographic
approach to QCD represents nothing but a five-dimensional language
for expressing some general expectations from the large-$N_c$
limit in QCD. This interpretation is valid only for holographic
actions quadratic in fields.

The action of SW
model is purely phenomenological and constructed with the aim of
getting the linear Regge spectrum. It is not known which 5D
dynamical model results in the background $e^{az^2}$ as a solution
of 5D Einstein equations. An "intermediate" dynamical model that
leads to such a background could look as follows,
\begin{equation}
\label{3bb} S=\int
d^4\!x\,dz\sqrt{g}\left(\partial_M\!\phi\partial^M\!\phi-m^2\phi^2+e^{\phi}\mathcal{L}\right),
\end{equation}
where the action is written in the AdS$_5$ space and $\mathcal{L}$
represents a Lagrangian density for the meson sector. The dilaton
coupling in the form $e^{\phi}\mathcal{L}$ is typical in the
string theory. If the 5D mass squared $m^2$ of the dilaton field
$\phi$ takes the minimal value permitted in the AdS$_5$ space,
$m^2R^2=-4$ (the Breitenlohner-Freedman stability
bound~\cite{freedman}) and one neglects backreaction from
$\mathcal{L}$, then the equation of motion for $\phi$ will have a
solution $\phi=az^2$. This solution leads to the $z$-dependent
background of SW model. One may speculate that the dilaton field
$\phi$ in~\eqref{3bb} mimics the non-perturbative gluodynamics. In
the large-$N_c$ limit, the gluon sector dominates. For instance,
the probability of creation of glueballs scales as
$\mathcal{O}(N_c^2)$ while that of mesons scales as
$\mathcal{O}(N_c)$~\cite{hoof}. We do not see an abundant
production of glueballs in the real world just because we cannot
create experimentally the corresponding gluon currents. But
theoretically the scaling $\mathcal{O}(N_c^2)$ of gluon sector
justifies neglecting backreaction from $\mathcal{L}$ in the first
approximation. In any case, the simplest SW model should be viewed
as an approximation in which a dynamical dilaton field is replaced
by a static dilaton background dictated by the phenomenology.

The use of the bottom-up models is much broader than just
modelling the hadron spectra --- it includes description of the
spontaneous chiral symmetry breaking in QCD, hadron formfactors,
QCD phase diagram, {\it etc}. The corresponding applications
usually contain certain extensions of simple bottom-up models that
go beyond the large-$N_c$ limit or have no clear connections with
the planar QCD. Establishing these connections is an open problem.

\end{document}